\documentclass[a4]{aa}
\usepackage{psfig}

\newcommand {\asec} {$^{\prime\prime}$}
\def\mic{$\,\mu\rm m$~}
\begin{document}
\thesaurus{10(07.03.01, 07.03.02, 13.09.5)}
\title{An ultra-deep ISOCAM observation through a cluster-lens
\thanks{Based on observations with ISO, an ESA project with instruments
funded by ESA Member States (especially the PI countries: France,
Germany, the Netherlands and the United Kingdom) with the participation
of ISAS and NASA}
}

\author{B. Altieri\inst{1}
\and L. Metcalfe\inst{1}
\and J.P. Kneib\inst{2}
\and B. McBreen\inst{3}
\and H. Aussel\inst{4}
\and A. Biviano\inst{5}
\and M. Delaney\inst{3}
\and D. Elbaz\inst{4}
\and K. Leech\inst{1}
\and L. L\'emonon\inst{4}
\and K. Okumura\inst{6}
\and R. Pell\'o\inst{2}
\and B. Schulz\inst{1}
}
\offprints{B. Altieri: baltieri@iso.vilspa.esa.es}
\institute{
ISO Data Centre, Astrophysics Division, Space Science Department of ESA, 
Villafranca del Castillo, P.O. Box 50727, E-28080 Madrid, Spain.
\and Observatoire Midi-Pyr\'ene\'es, 14 Av. E. Belin, 31400 Toulouse, France.
\and Physics Department, University College Dublin, Stillorgan Rd., Dublin 4, Ireland.
\and DSM/DAPNIA/SAp, CEA-Saclay, 91191 Gif-surYvette Cedex, France.
\and Osservatorio Astronomico di Trieste, via G.B. Tiepolo 11, I-34131 Trieste, Italy.
\and Institut d'Astrophysique Spatiale, B\^at. 121, Universit\'e Paris-Sud, F-91405 Orsay, France.
}
%
%\date{Draft \today}
\date{received October 1998; accepted asap}
\authorrunning{B. Altieri et al. 1998}
\titlerunning{An ultra-deep ISOCAM observation through a cluster-lens}
\maketitle
\markboth{B. Altieri et al. 1998}{Ultra-Deep MIR survey}

\begin{abstract}

We present results of ultra-deep ISOCAM observations through
a cluster-lens at 7$\mu$m and 15$\mu$m with 
the Infrared Space Observatory (ISO) satellite. 
These observations reveal a large number of luminous Mid-Infra\-red (MIR)
sources.  Cross-identification in the optical and Near-Infrared (NIR)
wavebands shows that about half of the 7\,$\mu$m
sources are cluster galaxies. The other 7$\,\mu$m and almost all
15\,$\mu$m sources are identified as lensed distant galaxies.
Thanks to the gravitational amplification they constitute the
faintest MIR detected sources, allowing us to extend the number
counts in both the 7 and 15\,\mic bands.
In particular, we find that the 15\,\mic counts have a steep
slope $\alpha_{15\mu{\rm m}} = -1.5 \pm 0.3$ and are large, 
with N$_{15\mu{\rm m}}(>\,30\mu{\rm Jy}) = 13 \pm 5$ arcmin$^{-2}$.
These numbers rule out non-evolutionary models and
favour very strong evolution.
Down to our counts limit, we found that the resolved 7\,\mic and 15\,\mic 
background radiation intensity is respectively (2$\pm$0.5)$\times$10$^{-9}$ 
and (5$\pm$1)$\times$10$^{-9}$ W\,m$^{-2}$\,sr$^{-1}$.

\end{abstract}
\section{Introduction}

Great progress in the understanding of physical properties of galaxies
has been achieved with Mid-Infrared (MIR) and Far Infrared (FIR)
observations using the ISO satellite (Kessler et al., 1996) and its ISOCAM
camera (Cesarsky et al., 1996).

Deep optical surveys, such as the Hubble Deep Field (Williams et al.,
1996) revealed a new population of distant sources at high-redshift
($z>2.5$) either using `drop-out' techniques (Steidel et al., 1996)
or `photometric redshift' ({\it e.g.} Lanzetta et al., 1996,
Connolly et al., 1997).
Yet optical surveys may miss a whole class of high-redshift
dust-enshrouded galaxies  (Blain \& Longair 1993).
In particular the progenitors of today's E/S0 galaxies at $z>$2 may emit
an important fraction of their total light in the NIR, MIR, FIR and
Sub-millimeter regions ({\it e.g.} Franceschini et al., 1994).
Indeed, galaxy formation models ({\it e.g.} Arimoto \& Yoshi 1987,
Guiderdoni et al., 1997)
predict that galaxies in their forming phase are heavily obscured
by abundant gas and dust inside the system.
These issues have motivated very deep ISOCAM observations of
blank fields (Rowan-Robinson et al., 1997;
Aussel et al., 1997; Aussel et al., 1998; D\'esert et al., 1998;
Taniguchi et al., 1997),
as well as Sub-millimeter observations 
(Smail, Ivison and Blain 1997; Blain 1997;
Barger et al., 1998.; Hughes et al., 1998).
 
The ISOCAM-HDF observations (Oliver et al., 1997; Aussel et al., 1998) 
showed that a non-evolving model at 15$\mu$m is ruled out at 3$\sigma$.
Analysing galaxy counts over a wide range of sensitivity, Elbaz et al., (1998) 
noted a change of slope in the counts observed at the mJy-level.
This variation cannot be explained by a simple evolutionary model, but
requires stronger star-formation activity for galaxies with $0.4<z<1$.

%Relatively deep ISOCAM observations
%have been conducted in the core of cluster lenses:
%A370 (Metcalfe et al., 1997, 1998), A2218 (Altieri et al., 1997), 
%A2390 (L\'emonon et al., 1998).
%Thanks to the gravitational amplification of the clusters,
%we reached fainter detection thresholds in a given observing time,
%and we were successful in unveiling the MIR emission
%of the giant arc in Abell 370 and other less distorted distant galaxies
%(Metcalfe et al., 1998).
Very deep ISOCAM observations have been conducted through gravitationally
lensing clusters as part of a large programme executed in the ISO Science
Operations Team guaranteed time (the ISO ARCS programme).
The clusters used were A370 (Metcalfe et al., 1997, 1999), A2218 (Altieri et 
al., 1998a; Metcalfe et al., 1999), MS2137, Cl2244 and A2390.
Further to a central goal of the ARCS programme, fainter detection thresholds 
were reached in a given observation time, thanks to gravitationnal 
amplification. We report here key results obtained for A2390.

By pushing ISOCAM to its ultimate limits with the help
of gravitational lensing, we performed 
ultra-deep observations through the core of probably
currently the best studied lensing cluster: Abell 2390 ($z=0.23$),
to obtain a magnified view of the background sky.
This both increases the sensitivity of our MIR maps and reduces the effects of 
source confusion.
These new observations supersede in area and depth
the previous ISOCAM observations of this cluster (L\'emonon et al., 1998).

In Section 2 the observations and data reduction techniques are discussed. 
Photometry, source identification and counts 
are presented in  Section 3. Results are discussed in Section 4.
Througout this paper, we used $\Omega=1$, $\lambda=0$ and 
$H_0=50 h_{50}$ km/s/Mpc.

\section{Observations \& Data Reduction}

\subsection{Observations}

The ESA guaranteed time observations reported here were allocated
one full ISO revolution (science window of 16 hours), 
between December 26 and 29, 1997.
We used the best observational strategy for detection of faint sources, based 
upon in-flight experience. 
In particular, the field was observed in 4 consecutive
revolutions in 4 blocks of 4 hours. The allocated time was equally divided into
the 2 broad-band ISOCAM filters 
LW2 (5-8.5$\mu$m), centered at 6.75$\mu$m, and LW3 (12-18$\mu$m), 
centered at 15 $\mu$m.

The observational strategy, that reaches the ultimate sensitivity
(Altieri et al., 1998b), is
performed by rastering the 32$\times$32-ISOCAM detector 
array in microscanning mode.
The pixel-field-of-view (PFOV) of 3\asec~ per pixel was chosen to obtain high 
spatial resolution at the expense of signal, but obtaining a better 
sampling of the PSF, 
crucial for source cross-identi\-fication.
The raster step size was 7\asec~ (2.33 pixels), the minimum value to
step out of the PSF FWHM in two consecutive pointings.
The size of rasters was 10$\times$10, so that, in the central part of each 
single raster, 100 different detector pixels sampled each sky pixel. Such 
redundancy was a key factor in ISOCAM deep observations.  
The final maps cover a field size of 2.6'$\times$2.6'.

\begin{table}
\begin{tabular}{lllccc}
\hline
Filter & Raster Size & Steps & N$_{exp}$ & T$_{int}$ & PFOV\\
\hline
LW2$\heartsuit$&4 - 10x10 & 7\asec  & 13 & 5s & 3\asec\\
LW2$\spadesuit$&1 - 6x6   & 10\asec & 30 & 5s & 3\asec\\
LW3$\heartsuit$&4 - 10x10 & 7\asec  & 13 & 5s & 3\asec\\
LW3$\spadesuit$&1 - 6x6   & 10\asec & 30 & 5s & 3\asec\\
\hline
\end{tabular}
\caption{Log of CAM Observation of the cluster-lens A2390.
$\heartsuit$ ESA GT data, $\spadesuit$ L\'emonon et al., (1998) data.}
\end{table}

\subsection{Data Reduction \& Calibration}

The Abell 2390 raw data of both the ARCS and the 
L\'emonon et al., (1998) programme were processed together following the steps:\\
{\sl i)} Dark subtraction using an IDL based package (Ott et al., 1997)
that uses a time-dependent dark correction (Biviano et al., 
1998).\\
{\sl ii)} Deglitching, flat-fielding and sky subtraction of each raster,
was performed using the PRETI Multi-resolution Median Transform 
techniques adapted for ISOCAM data analysis (Starck et al., 1997),
a multi-scale analysis that decomposes the different parts of the
signals.\\
{\sl iii)} Field distortion correction (Aussel et al., 1998) for sky
projection, with a final pixel size of 1\asec~.\\
{\sl iv)} The five (4+1) rasters were then stacked together using the
drizzling technique (Fruchter \& Hook 1998). The rasters were aligned 
using the centroids of the 4--5 brightest sources in each waveband.

\begin{figure}
\psfig{file=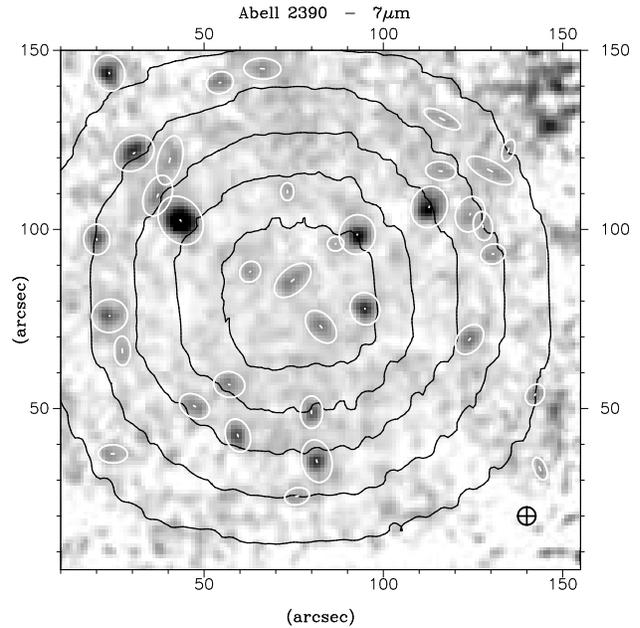,width=0.5\textwidth}
\caption{LW2 image of the A2390 cluster. The beam size (defined as 80\%
of encircled energy) is indicated in
the lower-righthand corner. Objects detected with the SExtractor software
are indicated by ellipses. Contours show the iso-exposure-time
of 5ks, 10ks, 15ks, 20ks and 25ks.}
\end{figure}

\begin{figure}
\psfig{file=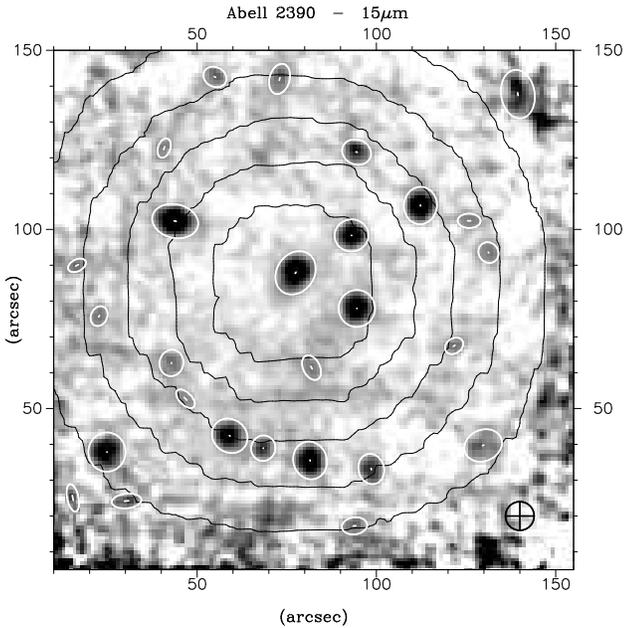,width=0.5\textwidth}
\caption{As Figure 1 but for the LW3 filter}
\end{figure}

The central parts of the stacked 7 and 15\,$\mu$m images are shown 
respectively in Figs. 1 \& 2.
The noise in the central part of the image 
is 0.1 $\mu$Jy arcsec$^{-2}$ in LW2 and 0.2 $\mu$Jy arcsec$^{-2}$ in LW3.

The calibration into $\mu$Jy was done using the
refined in-flight values from Blommaert et al. (1998).
We took into account
the transient behaviour of the ISOCAM pixel signals, which do not stabilize
at faint fluxes, but respond 
%to a change in illumination level
with transient drifts that depend both on the background level and the 
source intensity. 
We calibrated this response (in the 13 readouts per raster position) at 60\% in
LW2 and 80\% in LW3, at various levels by running the IAS transient correction
model (Abergel et al., 1996).

\section{Analysis \& results}

\subsection{Detection \& Photometry}

Source catalogs from our field in LW2 and LW3 were constructed using
the SExtractor package (Bertin \& Arnouts, 1996). 
%The detection algorithm uses
%the criterion that the surface brightness in 15 contiguous pixels
%exceeds a threshold (chosen as $\sim$1.5 $\sigma$ of the sky noise, Table 2),
%after subtracting a smooth background signal and convolving the image 
%by a 4x4 pixel top-hat filter.
The detection algorithm searches for 15 contiguous pixels, each having
a surface brightness exceeding a threshold
(chosen as $\sim$1.5 $\sigma$ of the sky noise, Table 2),
after subtracting a smooth background signal and convolving the image 
with a gaussian filter of the same width as the 7 and 15\,\mic image PSF.
We performed 7\asec~ aperture photometry. 
Aussel et al. (1998) showed that aperture photometry is very linear,
above 100\,$\mu$Jy
even though at the faintest fluxes the remnants of cosmic rays pollute the
sources with a positive bias.
The photometry was corrected for loss of flux in the PSF wings, by using
measurements of calibration stars under similar conditions (microscanning
\& rastering): 
60\% of the flux is found within a 7\asec~ diameter at 7\,\mic
and 48\% at 15\,\mic.

To assess the contribution of noise to our catalogs we ran the detection
algorithm on the negative fluctuations in the maps, concluding that we have
no false detections above 60 $\mu$Jy in the central 2'$\times$2'
area of the maps in both filter bands.

Finally, to determine the completeness of the sample we added a template
faint source (scaled-down version of a calibration star)
to the maps repeatedly at different positions in the map and
estimated the efficiency of detecting this source as a function of its flux
density. This provided a reliable estimate of the visibility of a
faint compact source in the maps. 
The estimated 80\% and 50\% completeness limits of
the catalogs derived from these simulations are listed in Table 2.
These figures refer to apparent source brightness and are similar to the 
deepest observations published to date at 7\,\mic on the Lockman Hole 
(Taniguchi et al., 1997,
faintest detections around 30$\mu$Jy) 
and at 15\,\mic on the HDF (faintest sources around 50 $\mu$Jy).
However, thanks to the gravitational magnification of a factor $\sim$2 to
10, the sources are intrisically the faintest
MIR sources detected to date.

\begin{table}
\begin{tabular}{llllll}
\hline
$\lambda$ & Threshold & N & N$_{neg}$  & S$_{80\%}$ & S$_{50\%}$\\
(\mic)    & $\mu$Jy/beam (3$\sigma$) & & & $\mu$Jy    & $\mu$Jy  \\
\hline
7         & 25 & 31 & 0 & 70  & 40\\
15        & 40 & 34 & 0 & 100 & 60\\
\hline
\end{tabular}
\caption{MIR source counts in the image plane, N: number of detected sources,
N$_{neg}$: number of detected sources on the inverted image}
\end{table}

\subsection{Cluster Contamination}

Thanks to our high-resolution images we have been able to
unambiguously identify almost all the MIR sources with counterparts in deep
NIR and optical (HST/WFPC2 and ground-based) images. 
The relative astrometric 
accuracy is found to be better than 1\asec~ in both filters.  In only a few 
cases we suspect that two sources are blended. 
There is one obvious case in the 7\,\mic map,
where the {\sl straight arc} (Pell\'o et al., 1991)
is blended with the nearby elliptical galaxy.

30 sources are detected at 7\,\mic. Half of them are
easily identified as cluster member galaxies
(Pell\'o et al., 1991, Leborgne et al., 1992, Abraham et al., 1996). 
The 5-8.5\,\mic emission of the cluster galaxies corresponds to 
4.5-6.9\,\mic restframe emission. For E/S0 galaxies it corresponds mostly to
the Rayleigh-Jeans tail of their old stellar population as in the Virgo cluster
(Boselli et al., 1998).
Two stars are identified, and at
least ten sources are lensed distant galaxies.
These lensed sources are all detected at 15\,\mic.

At 15\,\mic, 34 sources are detected in the central 2.25'$\times$\-2.25' field.
Only three sources are identified as cluster members: the cD galaxy 
(L\'emonon et al., 1998, Edge et al., 1998) and two star-forming galaxies.
Based upon spectroscopic or photometric redshifts, 
all the other sources are identified as faint lensed galaxies. 
Almost all sources for which we have spectroscopic redshifts are
background objects. 
Although, we can not rule out some of the targets being in the cluster,
the probability is very small.
The detection of almost exclusively background sources in the cluster images
demonstrates that at 15\,\mic the cluster-core becomes {\sl transparent}
(as in Sub-mm/UV bands).
Therefore the key feature is that the cluster-core acts as a natural 
gravitational telescope amplifying the flux 
of background sources, typically by a factor of 2. 

\subsection{Source counts}

By correcting for the lens magnification and surface dilution effects,
contamination by cluster galaxies,
and non-uniform sensitivity of our maps,
we can derive number counts at 15\,\mic to compare with {\it blank} sky  
counts (e.g. in the Hubble Deep field and 
Lockman Hole).
The 7\,\mic number counts are more difficult to derive due
to the larger contamination by the cluster and because of the small number
statistics.

The number density of sources is high with respect to the size of the FWHM
($\sim$ 6\asec~ diameter at 15\,\mic), but the PSF is well sampled on the final maps,
and its shape can be used to separate the sources;
only two 15\,\mic sources lie at the location
of pairs of suspected high-z galaxies.
Our counts are not significantly affected by confusion.

The occasional blending of the sources has not been taken into account, 
but the surface 
area occupied by bright sources is subtracted for the 
computation of the surface density of the fainter ones 
(ie. other faint sources could be hidden by brighter ones). 
We used the completeness of the detection at 15\,\mic given in \S~3.
This correction is negligible to the 7\,\mic counts and was not applied.

Due to the non-uniform sensitivity of our maps (because of observation
strategy) and the lensing effect,
the object density per flux bin was computed using gain-depen\-dant 
surface areas, and only the central 2'$\times$2' area 
was taken into account for the faintest fluxes.

A detailed lensing model of A2390 has been produced by Kneib et al., (1998).
The  lensing acts in two ways on the background population of galaxies:\\
{\it i)} an amplification of the source brightness,
typically by a factor of 2, but up to 10 near
the caustic lines.\\
{\it ii)} a surface dilation effect of the area probed, which 
itself depends on the redshift; the space dilation is stronger towards 
the centre (core of the cluster) and increases with source-plane redshift.

To estimate these factors we used: the spectroscopic redshift for
7 objects (Pell\'o et al., 1991, B\'ezecourt \& Soucail 1997), 
and for the rest we use the best redshift estimate obtained
with photometric redshift techniques (Pell\'o, private communication),
and/or lensing inversion techniques (Kneib et al., 1998). 
By analysing the case with all background galaxies at a mean redshift 
$\overline z=1.0$, we checked the dependence of the results on redshift 
uncertainties.
%We also ploted the counts with no-lensing correction, down to 60-70 $\mu$Jy.

\begin{figure}
\psfig{file=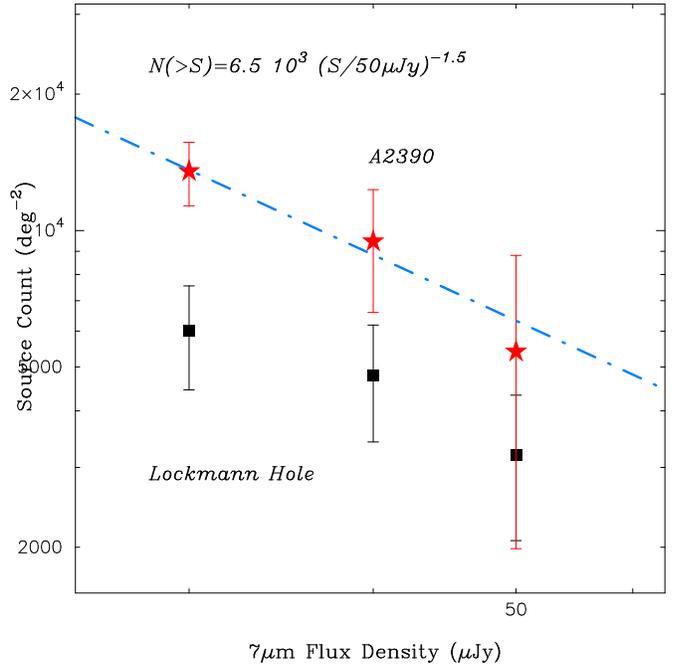,width=0.5\textwidth}
%\caption{7\,\mic lens-corrected counts, of identified field galaxies,
%compared to the Lockman Hole (dashed) counts (Taniguchi et al., 1997).}
\caption{7\,\mic lens-corrected counts, of identified field galaxies,
(filled stars) compared to the Lockman Hole counts (filled squares, Taniguchi et 
al., 1997).} 
\end{figure}

\begin{figure}
\psfig{file=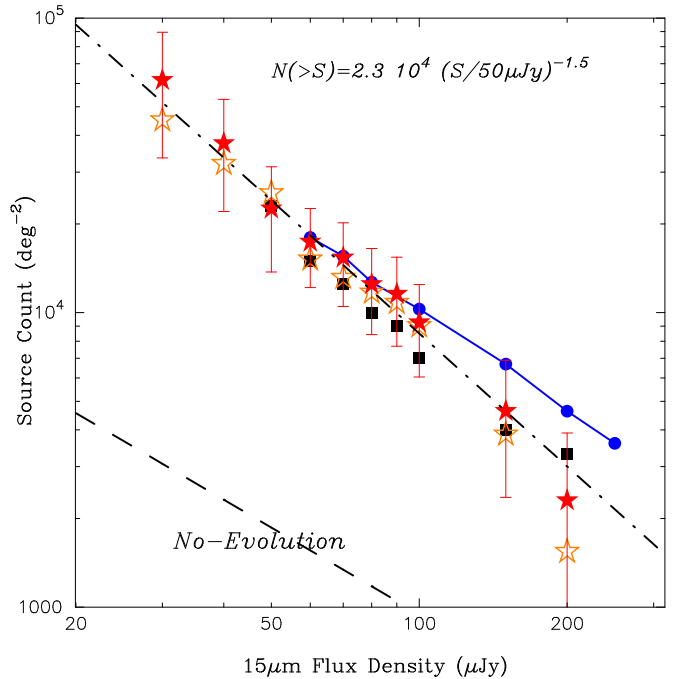,width=0.5\textwidth}
%\caption{15\,\mic lens-corrected counts corrected for incompleteness, 
%full model (solid),
%with all sources at z=1 ({\it diamonds} and dotted),
%without lensing correction ({\it triangles}) and
%compared to the HDF counts (Aussel et al., 1998)({\it squares} and dashed)
%and the no-evolution model (dotted line) from Franceschini et al., (1997).
%Note that the uncorrected counts are higher than the blank field counts.
%This is in good agreement with the steep slope measured which 
%produces a positive magnification bias. Error bars include both Poisson and
%systematic terms.
\caption{15\,\mic lens-corrected counts corrected for incompleteness, 
full model ({\it filled stars}),
with all sources at z=1 ({\it open stars}),
without lensing correction ({\it filled circles}) and
compared to the HDF counts (Aussel et al., 1998)({\it filled squares})
and the non-evolution model (dashed line) from Franceschini et al., (1997). 
Error bars include both Poisson and systematic terms.}
\end{figure}

\section{Discussion and Conclusion}

The source counts, corrected for cluster contamination and lensing
effects, in both the LW2 and LW3 bands, are presented in Figs.
3 and 4. We have used only 5$\sigma$ sources 
(i.e. $\sim$60 $\mu$Jy before lensing amplification 
correction). According to Hogg \& Turner (1998),
this is sufficient to avoid the positive flux-estimate Eddington bias 
which occurs for faint source counts.

At 7\,\mic, we find that the source density is greater than that of
Taniguchi et al., (1997) by a factor of 2. We suggest that their observations 
are incomplete below 50\,$\mu$Jy, whereas our 7\,\mic map is 80\% complete 
down to 30\,$\mu$Jy. 

At 15\,\mic the number counts are compatible with the results of 
Aussel et al. (1998) in the HDF.
However, with the help of gravitational lensing
we are able to extend the counts in both bands down to 30$\mu$Jy.
Putting all background sources at $z=1$ only slightly decreases the
faintest counts, at the faintest end,
because a few suspectedly high-z faint sources would be less amplified.
This shows the small dependency of our derived counts on the models. 
The very small positive magnification bias in the lensing-uncorrected counts
is consistent with the measured slope of the source counts,
while the general shift of the points towards higher flux follows
from the loss of the lensing gain.

We find a total number density
N$_{7}$($>$\,30$\mu$Jy) = 3.5 $\pm 1$ arcmin$^{-2}$ at 7\,\mic,
and $\alpha_{15} = -1.5 \pm 0.3$, and
N$_{15}$($>$\,30$\mu$Jy) = 13 $\pm 5$ arcmin$^{-2}$ at 15\,\mic.

The 15 $\mu$m counts show a steadily increasing excess 
(by more than a factor of 10) with respect to the
prediction of a no-evolution model (dotted line, Franceschini et al., 1997).
This confirms the steeper count slope below 1~mJy found on the 
Lockman Hole (Elbaz et al., 1998) and is in good agreement with 
the ISO HDF counts (Aussel et al., 1998).
The counts are a factor 2-3 higher than the boundaries of
the counts coming from an early analysis of the background 
fluctuations in the ISOCAM 15\,\mic map of the HDF (Oliver et al., 1997).
In particular, we do not seem to detect any sign of
flattening of the counts at the faintest levels,
as expected from evolutionary models
(Franceschini et al., 1997, Oliver et al., 1997).
The slope stays close to -1.5$\pm 0.3$ down to 30 $\mu$Jy.
This source density at faint levels favours extreme evolution models, 
needed to fit the counts at brighter fluxes, as shown by Elbaz et al. (1998).

Integrating the number counts over the whole flux range and extrapolating the
counts we find that respectively $(5 \pm 1) \times 10^{-9}$ and
$(2.0\pm 0.5)\times 10^{-9}$ W\,m$^{-2}$\,sr$^{-1}$ is emitted at 15\,\mic and 
7\,\mic.  This implies that a larger fraction of the UV/optical background 
is re-radiated in the MIR than in the local universe.

The absolute astrometry of MIR sources is difficult in deep surveys 
because of the lack of any obvious optical-NIR counterparts.
In the A2390 field, unambiguous cross-identification of more than 90\%
of the sources was possible thanks to a large density of
sources and a good sampling of the PSF.
A number of these sources are well correlated with faint galaxies in the 
visible, some of them having very red colors in the NIR.
%however, none of them can be distinguished
%from other field galaxies based on optical colours only. 

15\,\mic ISOCAM deep imaging is a good way 
to select star-forming galaxies and dusty AGNs
which are not easy to identify in UV/optical surveys.
A more detailed analysis of the SED of these MIR detected galaxies
will be necessary to  determine the nature of these sources
and give an estimate of their SFR.
Our observations confirm
that abundant star formation activity occurs 
in very dusty environments at $z\sim 1$. 
Great caution must therefore be taken to infer global star
formation activity based only on UV-continuum or optical 
luminosities of high-z galaxies.
The global star formation history can be traced fully only if 
the effect of dust is taken into account in a consistent way. 
({\sl e.g.} Blain et al., 1998). Further, detailed analysis of our selected
sample is needed to unveil the nature of these MIR galaxies.

\acknowledgements {\scriptsize
JPK acknowledges support from CNRS/INSU.
Many thanks to A. Blain and  I. Smail
for useful discussions and comments.
The ISOCAM data presented in this paper was analysed using ``CIA",
a joint development by the ESA Astrophysics Division and the ISOCAM
Consortium. The ISOCAM Consortium is led by the ISOCAM PI, C. Cesarsky,
Direction des Sciences de la Matiere, C.E.A., France.
}

\end{document}